# Improving Photometric Redshifts by Merging Probability Density Functions from Template-Based and Machine Learning Algorithms

Ishaq Y. K. ALSHUAILI (https://orcid.org/0000-0003-4064-9253)[a], John Y. H. SOO (https://orcid.org/0000-0001-5328-0892)[a,*], Mohd. Zubir MAT JAFRI (https://orcid.org/0000-0001-9313-7548)[a] and Yasmin RAFID[a]

[a] School of Physics, Universiti Sains Malaysia, 11800 USM, Pulau Pinang, Malaysia

*e-mail: johnsooyh@usm.my

Abstract—This study aims to improve the photometric redshifts (photo-zs) of galaxies by integrating two contemporary methods: template-fitting and machine learning. Finding the synergy between these two methods was not a high priority in the past, but now that our computer processing power and observational accuracy have increased, we deem it worth investigating. We compared two methods to improve galaxy photometric redshift estimations by using the algorithms ANNz2 and BPz on different photometric and spectroscopic samples from the Sloan Digital Sky Survey (SDSS). We find that the photometric redshift performance of ANNz2 (machine learning) is better than that of BPz (galactic templates), and with the utilization of the merging technique we introduced, we see that there is an improvement in photo-z when the two strategies are consolidated, providing improvements in $\sigma_{\text{RMS}}$ and $\sigma_{68}$ up to [0.0265 & 0.0222] in LRG sample and [0.0471 & 0.0471] in the Stripe-82 sample. This simple demonstration can be used for photo-zs of galaxies in fainter and deeper sky surveys, and future work is required to prove its viability in these samples.

## 1. INTRODUCTION

The measurement of galaxy redshifts represents an important topic to be studied, since it is required in many cosmological research work as it is also essentially adding a third, radial dimension to cosmological investigations. They make it possible to examine phenomena as a function of time and distance, as well as to identify structure formations like galaxy clusters, measuring distance-dependent quantities such as luminosities and masses. Redshifts are also necessary to separate large-scale structures and galaxies along the line of sight. Even today, the issue of obtaining photometric redshift (photo-z) estimations that are precise enough to suit the needs of cosmology and galaxy evolution investigations drives active development of photometric methods and photo-z algorithms.

In the past, redshifts of galaxies have been calculated in various ways. One of the ways is to look at properties in redshifted galaxy spectra and compare them to the known rest-frame spectra of molecules and atoms on Earth using spectroscopic redshifts, often denoted as spec-zs or $z_{\text{spec}}$. As a result, spectroscopic galaxy studies have been contributed towards the understanding of the origin, composition, and evolution of the Universe. The current set of spec-zs is insufficient for most current cosmological studies, mainly because spectroscopy, unlike photometry, is a time-consuming and expensive technique, and spectroscopic measurements are constantly limited by currently available technology and optics (Soo. 2018). As such, in order to produce redshifts ideally for all objects in large galaxy samples, the concept of photometric redshifts (photo-zs) was born.

The photometric redshift technique described in the literature can be classified into two broad categories: the empirical training set method and the fitting of spectral energy distributions (SED) by synthetic or empirical template spectra. The first approach is also known as the machine learning method, an empirical relationship between magnitudes and redshifts is derived using a subsample of objects (the training set) in which both the redshifts and photometry are available (Connolly et al. 1995). A slightly modified version of this method was used by (Wang et al.1998) to derive redshifts in the Hubble Deep Field (HDF–N) by the means of a linear function of colours. In the SED-fitting approach, a spectral library is used to compute the colours of various types of sources at any plausible redshift, and a matching technique is applied to obtain the "best-fitting" redshift. This technique has been used extensively in deep cosmological surveys (Vanzella et al. 2003).

Although most methods proposed to improve the estimates of galaxy redshifts have showed that machine learning methods have a clear performance advantage, however, both template and machine learning methods have their own advantages that we can benefit from. For example, template fitting methods can recognise bad or improbable template matches, and wide or multipeaked posterior redshift PDFs, while machine learning methods can signal extrapolation, large deviations among neighbours, or sparse training set coverage (Beck et al. 2017). Thus, taking advantage of the synergy of these two methods can be an opportunity to achieve more accurate results when combined (Newman et. al 2022).

In this work, we present how we can improve photometric redshift of galaxies via exploring the synergy of two methods: template and machine learning, by merging the photo-z probability density function (PDFs) resulting from the two methods in different ways to improve the overall galaxy redshift estimation. We use galaxy data from the Sloan Digital Sky Survey (SDSS) and we use two photo-z algorithms, ANNz2 and BPz to represent the machine learning and template fitting methods, respectively, on two different galaxy samples.

The proposed methodology seeks to improve and minimise the typical deviations of point estimates from their true values, reduce the frequency of outliers, provide a conservative range of possibilities for the redshift of an object and create PDFs that do better statistical definition. These objectives are in line with the main goals of current photometric redshift research (Newman et. al 2022). It can be said that the methods used in this study has its novelty within other previous methods and significantly improve the accuracy of the redshifts of galaxies.

The structure of this paper is as follows: in Section 2 we describe the photometric and spectroscopic data used. Section 3 describes the photometric redshift algorithms used in this work, while Section 4 describes the methods we used to merge the PDFs. A discussion of the main results in the paper can be found in Section 5. Finally, we present our conclusions in Section 6.

## 2. PHOTOMETRIC DATA

### 2.1. Sloan Digital Sky Survey (SDSS)

With deep multi-colour photographs of one-third of the sky and spectra for more than three million celestial objects, the Sloan Digital Sky Survey (SDSS) has developed the most detailed three-dimensional map of the Universe ever constructed. Through the 2.5 m-telescope at the Apache Point

Observatory, it is equipped with a large-format mosaic CCD camera to image the sky in five optical bands (York et al. 2000). SDSS has now completed four phases of operations with a fifth ongoing. Since 2017, SDSS has had a dual hemisphere view of the sky (Abdurro'uf et al. 2022). SDSS's latest data release (DR17) includes the Apache Point Observatory Galactic Evolution Experiment 2 (APOGEE-2) survey, which publicly distributes infrared spectra of nearly 650 000 stars. DR17 also offers 25 new or updated Value-Added Catalogues (VACs) in addition to the primary data sets.

SDSS aims at producing a large homogeneous data sample of the northern sky with high accuracy multicolour photometry and accurate astrometry. To date, SDSS has obtained photometry for more than 300 million objects with a median seeing of $1.4''$ in the $r$-band. The reader is referred to Gunn et al. (1998) for more information on the telescopes and cameras used by SDSS. In this work, we use two different data sets from SDSS, the first sample is the Luminous Red Galaxy Sample (with 13920 galaxies) and it is suited to study large-scale structures, clusters and the evolution of giant elliptical galaxies. The second sample is the Stripe-82 (with 9594 galaxies). We chose to work with objects in the Stripe-82 region due to the abundance of spectroscopic redshifts in this region.

### 2.2. The Luminous Red Galaxy (LRG) Sample

Large red elliptical galaxies have been discovered to exist in massive dark matter halos and cluster strongly (Soo et al. 2018). In most galaxy clusters and groups, the brightest galaxies have a narrow range of hue and intrinsic luminosity. These galaxies are known as "bright red galaxies" because they are the most massive, luminous, and reddest (in the rest-frame colour) of all galaxies. These luminous red galaxies (LRGs) are good tracers of the universe's large-scale structure due to their brilliant intrinsic luminosities and strong clustering. Several research including the SDSS have already used LRGs to explore large-scale structure (York et al. 2000, Eisenstein et al. 2011). LRGs are made up of a homogeneous, roughly volume-limited selection of objects with the reddest hues in the rest frame and are chosen using a variant of the photometric redshift technique. Cuts in the $g - r, r - i, r$ colour-colour-magnitude cube are used to produce the SDSS LRG sample.

In this work, we used approximately 13 920 galaxies within the redshift range of $0.3 < z < 0.8$.

### 2.3. SDSS Stripe-82 Sample

Stripe-82 is a 2.5-degree wide stripe in the Southern Galactic Cap ($-50° < \alpha < 60°, -1.25° < \delta < 1.25°$) that spans all five SDSS bands and covers a total size of $275 \text{ deg}^2$. The SDSS observed it numerous times as part of the Legacy Survey between 1998 and 2004 in 82 runs (94 to 5052) under photometric conditions with typically good viewing and low sky background. This was first done to allow the repetitive imaging scans to be combined in order to achieve fainter magnitudes, around 2 mag fainters than the single SDSS scans (Annis et al. 2011). The first version of co-added images made from the Stripe-82 images were released in the SDSS Data Release 7 (Abazajian et al. 2009).

In this paper, we used 9594 galaxies with a redshift range of $0.0 < z < 1.2$.

## 3. PHOTOMETRIC REDSHIFT ALGORITHMS

The basic photometric redshift approach involves using the colours of a galaxy through a variety of medium- or broadband filters as a rough approximation of the galaxy's SED to determine its redshift and spectral type (Firth et. al 2003). Previously, such distances could only be determined for small samples of objects due to the cosmological expansion's displacement of spectral features, but spectroscopy, which is time-consuming and expensive, could not be used effectively on very faint sources or large samples of galaxies. Thus, the alternative methodology of photo-z estimation methods was developed as a result of this. The photo-z methodology was originally presented by Baum (1962) and better codified by Butchins (1981) and in the key studies of Connolly et al. (1995) and Brescia et al. (2021).

In this work, we present how we can improve photometric redshift of galaxies via exploring the synergy of two methods: template and machine learning, by compare probability density function (PDFs) of the two methods and find different ways to improve galaxy redshift estimations by analysing galaxy data from the Sloan Digital Sky Survey (SDSS) and using ANNz2 and BPz codes on different photometric and spectroscopic samples.

### 3.1 ANNz2

ANNz2 (Sadeh et al. 2016) is one of the latest artificial neural networks (ANN) photo-z algorithms available in the literature. ANNz2 is a significant improvement from its predecessor ANNz (Sadeh et al. 2016) in that it includes boosted decision trees (BDTs) and k-nearest neighbours (KNN) in addition to ANNs in its algorithm. ANNz2 utilises the Toolkit for Multivariate Data Analysis (TMVA) package (Sadeh et al. 2016) to train with various machine learning methods and generate photo-zs based on a weighted average of their results. ANNz2 is able to weight the training data to imitate the target set, and by providing a correction factor to the training objects based on the input parameters, these weights may subsequently be transmitted throughout the training and assessment process. ANNz2 uses a high level of task automation with python scripts, eliminating the need for the user to define specific machine learning method (MLM) attributes which allows for a significant improvement in photo-z quality without requiring considerable user participation in the training process. By folding chosen individual MLM findings with their uncertainty estimates, the randomized regression mode of ANNz2 enables the generation of probability distribution functions (PDFs) of the calculated photo-zs. These PDFs should not be interpreted as genuine error distributions in regard to the true redshift (which is unknown), but as a quantification of the photo-z derivation method's uncertainties (Bilicki et al. 2018).

In this work, we chose to use ANNz2 because it is capable of doing regression estimation of single-value photo-z solutions as well as PDFs. We optimise ANNz2 by employing the $N$:2$N$:2$N$:1 architecture that is the default option for ANNz, which enables us to obtain better outcomes and improve the value of photo-zs.

### 3.2 BPz

Bayesian Photometric Redshift (BPz, Benítez 2000) is a freely available template-based photo-z code with a sophisticated Bayesian apparent magnitude/type prior formulation. It was one of the main codes used to produce redshift distributions estimates through stacking of the individual redshifts for CFHT Legacy Survey (CFHTLS, Ilbert et al. 2006) and Dark Energy Survey (DES, Abbott et al. 2005). By modelling accurately their spectral evolution, BPz can provide estimates for redshifts from our knowledge of galaxy spectra, up to high redshifts if needed, obviating the requirement for expensive and biased spectroscopic source measurements for training sets.

To choose the best template sets for BPz in this work, we have compared the results of several template sets (see **Table 1**), and we find that the root-mean-square error ($\sigma_{\text{RMS}}$) obtained when using the Brown templates gave us the best results for both the LRG ($0.3 < z < 0.8$) and Stripe-82 ($0.1 < z < 1.2$) samples, respectively. In all tests, we chose a step size of 0.002 for the PDFs and an interpolation of 2 between the templates. **Table 2** shows the results of our optimisation process, and based on the results, we chose to use Brown templates in all subsequent runs of BPz.

## 4. MERGING PHOTO-Z PDFS

In this work, we explore the synergy between ANNz2 and BPz by merging their PDFs to see if we can improve the overall photo-z performance. We attempted three different methods to merge the PDFs. the first method is to combine the PDFs of the two algorithms directly and averaging them, namely

$$\text{PDF}_{\text{avg}} = \frac{\text{PDF}_1 + \text{PDF}_2}{2}, \tag{1}$$

where $\text{PDF}_1$ represents the PDF from ANNz2, while $\text{PDF}_2$ represents the PDF from BPz.

The second method uses a weighted sum between the two PDFs, it relies on finding a value $\alpha$ such that

$$\text{PDF}_{\text{wgt}} = \alpha \text{PDF}_1 + \alpha \text{PDF}_2, \tag{2}$$

where $\alpha$ is calculated based on the performance ratio between BPz and ANNz2. In our calibration process, we find that 57.1% of the time, ANNz2 outperforms BPz in the LRG sample, while the number is 63% in the Stripe-82 sample. Thus, we set $\alpha = 0.571$ for the LRG sample and $\alpha = 0.630$ for the Stripe-82 sample.

In the final method, we combine both PDFs by multiplying both of them together, mathematically this gives

$$\text{PDF}_{\text{mtp}} = \frac{\text{PDF}_1 \odot \text{PDF}_2}{\sum_j \text{PDF}_1 \odot \text{PDF}_2}, \tag{3}$$

where $\odot$ refers to the operator for element-wise multiplication, and the area of the new PDF normalised to 1.

Using these three methods, we create new PDFs and therefore creating new photo-zs denoted by $z_{\text{avg}}$, $z_{\text{wgt}}$ and $z_{\text{mtp}}$. Each of these photo-zs can be generated from the peak or the mean of the PDFs, respectively, and both will be studied and analysed in this work. We will use three performance metrics, the $\sigma_{\text{RMS}}$ (root-mean-square error), the $\sigma_{68}$ (68$^{\text{th}}$ percentile error) and the outlier rate fraction ($\eta_{\text{out}}$) to determine the magnitude of improvement in the combined photo-zs with respect to the individual photo-zs of ANNz2 and BPz. We use the same definitions of $\sigma_{\text{RMS}}$, $\sigma_{68}$ and $\eta_{\text{out}}$ as used in Soo et al. (2018).

## 5. RESULTS

In this work, we computed the photo-zs using the machine learning algorithm ANNz2 and the template-based algorithm BPz, training / optimising each code on the same data samples (LRG and Stripe-82), and we used all five $ugriz$ magnitudes as inputs. The results of the individual algorithms are shown in **Table 3** and visualised in **Fig. 1**, where we see that ANNz2 performed better than BPz in both cases.

*5.1 Photo-zs on the LRG Sample*

For the LRG sample, we find that the performance metric values have improved dramatically with the application of the methods listed in **Section 4** with varying degrees of optimisation. This is more evident when applying the method of $PDF_{mtp}$ in the LRG sample, and we can see this clearly in the values of $\sigma_{RMS}$ and $\sigma_{68}$ as shown in **Table 4** and visualised in **Fig. 2**. From **Fig. 2**, when comparing $PDF_{mtp}$ to the performance of other techniques, we see that the scatter at the high redshift regime originally seen in BPz has vanished using this combination method.

We also randomly selected 8 galaxies at different redshift intervals, and plot the photo-z PDFs of ANNz2, BPz and each combination method in **Fig. 3**. From **Fig. 3**, we find that the performance of the averaged and weighted PDFs is close to the performance of ANNz2, but however the multiplied PDF showed a better performance as the peaks of the $PDF_{mtp}$ are closer to their respective spectroscopic redshift values (black line), which in turn produces an overall lower value of $\sigma_{68}$.

When comparing our results to a previous work on SDSS LRGs, we find that our $\sigma_{RMS}$ value (0.028) for galaxy photo-zs is significantly lower than those of Abdalla et al. (2011), which used the older version of ANNz (at $\sigma_{RMS} = 0.058$) and several other template codes. Comparing to a more recent results, we see that our $\sigma_{RMS}$ is higher than that of Meshcheryakov et al. (2015), which had a value of $\sigma_{RMS} = 0.011$. We note that this discrepancy is mainly due to different sample selections, as they have used the BOSS LOWZ sample which has a lower redshift range of $z < 0.45$ and an updated spectroscopic redshift sample with lower errors. Thus, we are confident that our results are competitive with respect to our sample selection, and we will attempt to test our architecture on other well-known samples in the future.

*5.2 Photo-zs on the Stripe-82 Sample*

The improvement in photo-z performance is also observed in the Stripe-82 sample, where the values of $\sigma_{RMS}$ and $\sigma_{68}$ are shown in **Table 4** and visualised in **Fig. 4**. Compared to the performance of ANNz2, it is evident that the multiplied PDFs performed well within the central redshift range, while the scatter at the high redshift region is significantly reduced. Scattering is also reduced when compared to the performance of BPz and this is due to the help provided by ANNz2.

Also, we note that the great performance of the multiplied PDF method can be seen prominently in **Fig. 5**. Taking the second and seventh rows as examples, we see that BPz underperformed here due to the existence of multiple peaks, while ANNz2 has managed to get relatively good results. It is shown that the multiplied PDF produced from ANNz2 and BPz successfully suppresses one of the peaks, therefore improving the overall photo-z of those galaxies significantly, in contrast to the degradation caused by the averaged and weighted PDFs. These results provide us a good indication that the two algorithms ANNz2 and BPz worked well together when using the multiplied PDF method, and despite being a very simple method, this showed great potential in improving photo-z values when combining several photo-z algorithms together.

## 6. CONCLUSION AND FUTURE WORK

Template-fitting and machine learning methods both produce good photometric redshifts when applied to different galaxy samples, as each method relies on different methodologies to produce the desired output. While BPz relies greatly on the type of SED template set used, ANNz2 heavily relies on large and reliable training sample, each restricted to the broadband filters and redshift ranges they are worked on. Each of these methods has its own advantages and disadvantages, thus by taking advantage of the advantages of each method and creating synergies between them, this will contribute to finding better and more accurate photo-zs.

This study has made use of the three different statistical methods, namely by creating averaged, weighted and multiplied PDFs where each effectively contributed to improve the resulting photo-zs by taking advantage of this synergy. Through this simple test of results using the LRG and Stripe-82 samples, we see that a multiplied PDF generated from ANNz2 and BPz has produced better photo-zs than what each algorithm could produce separately. The multiplied PDF showed an improvement in $\sigma_{RMS}$ of 8.3% and 2.9% as compared to ANNz2 in the LRG and Stripe-82 samples, respectively. These encouraging results show the potential for this method to be used in larger and fainter surveys like the Legacy Survey of Time and Space (LSST, Ivezic et al. 2019) and the Physics of the Accelerating Universe Survey (PAUS, Padilla et al. 2019), both of which utilises group effort to produce reliable photometric redshifts.

Therefore, we believe that this work could be further experimented on different galaxy samples and different photo-z algorithms to test the effectiveness of this method. We also seek to explore methodologies to smoothen PDFs (like in the case of ANNz2 as shown in this work), and test if this affects the overall photo-z results. We believe there is still room for improvement in this research area, and it is hoped that we intend to explore other methodologies related to synergising different photo-z algorithms.

## ACKNOWLEDGEMENT

Funding for the SDSS and SDSS-II has been provided by the Alfred P. Sloan Foundation, the Participating Institutions, the National Science Foundation, the U.S. Department of Energy, the National Aeronautics and Space Administration, the Japanese Monbukagakusho, the Max Planck Society, and the Higher Education Funding Council for England. The SDSS Web Site is http://www.sdss.org/.

The SDSS is managed by the Astrophysical Research Consortium for the Participating Institutions. The Participating Institutions are the American Museum of Natural History, Astrophysical Institute Potsdam, University of Basel, University of Cambridge, Case Western Reserve University, University of Chicago, Drexel University, Fermilab, the Institute for Advanced Study, the Japan Participation Group, Johns Hopkins University, the Joint Institute for Nuclear Astrophysics, the Kavli Institute for Particle Astrophysics and Cosmology, the Korean Scientist Group, the Chinese Academy of Sciences (LAMOST), Los Alamos National Laboratory, the Max-Planck-Institute for Astronomy (MPIA), the Max-Planck-Institute for Astrophysics (MPA), New Mexico State University, Ohio State University, University of Pittsburgh, University of Portsmouth, Princeton University, the United States Naval Observatory, and the University of Washington.

FUNDING

JYHS and YR acknowledge financial support by Universiti Sains Malaysia (USM) through the Short-Term Research Grant (Geran Penyelidikan Jangka Pendek) with account number 304/PFIZIK/6315395.

CONFLICT OF INTEREST

The authors declare that they have no conflicts of interest.

TABLES

**Table 1**. The 5 template sets we used to optimise BPz in this work, they are listed by name, number of templates in the sets and their and their citation sources.

| Template Set Name | No. of Templates | Citation |
|---|---|---|
| Brown | 129 | Brown et al. (2014) |
| LePhare | 66 | Coleman et al. (1980), Kinney et al. (1996) |
| COSMOS | 31 | Bruzual & Charlot (2003), Polletta et al. (2007) |
| CWW-Kinney | 10 | Coleman et al. (1980), Kinney et al. (1996) |
| CWW | 8 | Coleman et al. (1980) |

**Table 2**. $\sigma_{RMS}$ and $\sigma_{68}$ of the photo-zs for both the MGS and Stripe-82 samples during optimisation, showing the different number of interpolations between templates in BPz using the template sets listed in **Table 1**. The lowest values are highlighted in green.

| Template Set | Interpolations | LRG | | Stripe-82 | |
|---|---|---|---|---|---|
| | | $\sigma_{RMS}$ | $\sigma_{68}$ | $\sigma_{RMS}$ | $\sigma_{68}$ |
| Brown | 0 | 0.0309 | 0.0247 | 0.0673 | 0.0331 |
| | 1 | 0.0304 | 0.0244 | 0.0680 | 0.0317 |
| | 2 | 0.0302 | 0.0242 | 0.0684 | 0.0307 |
| | 3 | 0.0302 | 0.0242 | 0.0680 | 0.0305 |
| LePhare | 0 | 0.0372 | 0.0304 | 0.0634 | 0.0340 |
| | 1 | 0.0371 | 0.0302 | 0.0633 | 0.0337 |
| | 2 | 0.0371 | 0.0302 | 0.0633 | 0.0336 |
| | 3 | 0.0371 | 0.0301 | 0.0633 | 0.0336 |
| COSMOS | 0 | 0.0343 | 0.026 | 0.0672 | 0.0336 |
| | 1 | 0.0343 | 0.0258 | 0.0672 | 0.0335 |
| | 2 | 0.0342 | 0.0258 | 0.0673 | 0.0335 |
| | 3 | 0.0341 | 0.0258 | 0.0673 | 0.0335 |
| CWW-Kinney | 0 | 0.0482 | 0.0268 | 0.0822 | 0.0636 |
| | 1 | 0.0372 | 0.0304 | 0.0765 | 0.0585 |
| | 2 | 0.0463 | 0.0437 | 0.0740 | 0.0553 |
| | 3 | 0.0453 | 0.0435 | 0.0733 | 0.0532 |
| CWW | 0 | 0.0361 | 0.0266 | 0.0734 | 0.0531 |
| | 1 | 0.0353 | 0.0277 | 0.0679 | 0.0469 |
| | 2 | 0.0350 | 0.0274 | 0.0665 | 0.0446 |
| | 3 | 0.0350 | 0.0267 | 0.0661 | 0.0440 |

**Table 3**. Comparison of $\sigma_{\text{RMS}}$, $\sigma_{68}$ and $\eta$ outlier fraction for ANNz2 and BPz on the LRG and Stripe-82 samples.

| Galaxy Sample | Redshift Range | Algorithm | $\sigma_{\text{RMS}}$ | $\sigma_{68}$ | $\eta_{\text{out}}$ |
|---|---|---|---|---|---|
| LRG | $0.3 < z < 0.8$ | ANNz2 | 0.0289 | 0.0252 | 0.0647 |
|  |  | BPz | 0.0302 | 0.0242 | 0.1221 |
| Stripe-82 | $0.1 < z < 1.2$ | ANNz2 | 0.0485 | 0.0221 | 1.8657 |
|  |  | BPz | 0.7015 | 0.0350 | 4.3152 |

**Table 4**. Performance of the photo-z as shown in the root-mean-square ($\sigma_{\text{RMS}}$), 68th percentile ($\sigma_{68}$) errors and outlier fraction ($\eta_{\text{out}}$) when applying the three PDF merging methods for the LRG and Stripe-82 samples. The lowest values are shown in green, and the results of ANNz2 and BPz are shown for reference.

| Method | Photo-zs Generated From the PDFs | LRG | | | Stripe-82 | | |
|---|---|---|---|---|---|---|---|
| | | $\sigma_{\text{RMS}}$ | $\sigma_{68}$ | $\eta_{\text{out}}$ | $\sigma_{\text{RMS}}$ | $\sigma_{68}$ | $\eta_{\text{out}}$ |
| $\text{PDF}_{\text{avg}}$ | Peak | 0.0268 | 0.0226 | 0.0718 | 0.0499 | 0.0227 | 2.0533 |
| | Mean | 0.0269 | 0.0228 | 0.0647 | 0.0490 | 0.0238 | 2.0221 |
| $\text{PDF}_{\text{wgt}}$ | Peak | 0.0268 | 0.0226 | 0.0718 | 0.0489 | 0.0228 | 1.9179 |
| | Mean | 0.0268 | 0.0226 | 0.0647 | 0.0489 | 0.0228 | 1.9074 |
| $\text{PDF}_{\text{mtp}}$ | Peak | 0.0268 | 0.0226 | 0.0718 | 0.0504 | 0.0227 | 2.0846 |
| | Mean | 0.0265 | 0.0222 | 0.0718 | 0.0471 | 0.0216 | 1.7510 |
| ANNz2 | - | 0.0289 | 0.0252 | 0.0647 | 0.0485 | 0.0222 | 1.8657 |
| BPz | - | 0.0302 | 0.0242 | 0.1221 | 0.0504 | 0.0227 | 4.3152 |

FIGURE CAPTIONS

**Fig. 1**. Photo-z vs. spec-z plots of ANNz2 (green) and BPz (blue) on the LRG (left) and Stripe-82 (right) samples, respectively.

**Fig. 2**. Photo-z vs. spec-z plots comparing the performance of ANNz2 (top left) and BPz (bottom left) with the methods we used to improve the photo-zs in the LRG sample. It can be seen that the multiple method gives the best performance as compared to the rest.

**Fig. 3**. An illustration of 8 randomly selected photo-z PDFs from the LRG sample produced by ANNz2 (green), BPz (red), and when combined using the three methods: averaged (left), weighted (middle) and multiplied (right). The resultant PDFs (blue) showed that all three methods produced similar results.

**Fig. 4**. Photo-z vs. spec-z plots comparing the performance of ANNz2 (top left) and BPz (bottom left) with the methods we used to improve the photo-zs in the Stripe-82 sample. It can be seen that the multiplied PDF gave a better performance when compared to the rest of the methods, as well as those of ANNz2 and BPz.

**Fig. 5**. An illustration of 8 randomly selected photo-z PDFs from the Stripe-82 sample produced by ANNz2 (green), BPz (red), and when combined using the three methods: averaged (left), weighted

(middle) and multiplied (right). Row 7 shows how the multiplied PDF outperforms the other methods by suppressing the anomalous photo-z peak of BPz, successfully giving a photo-z closer to the original spec-z of the object.

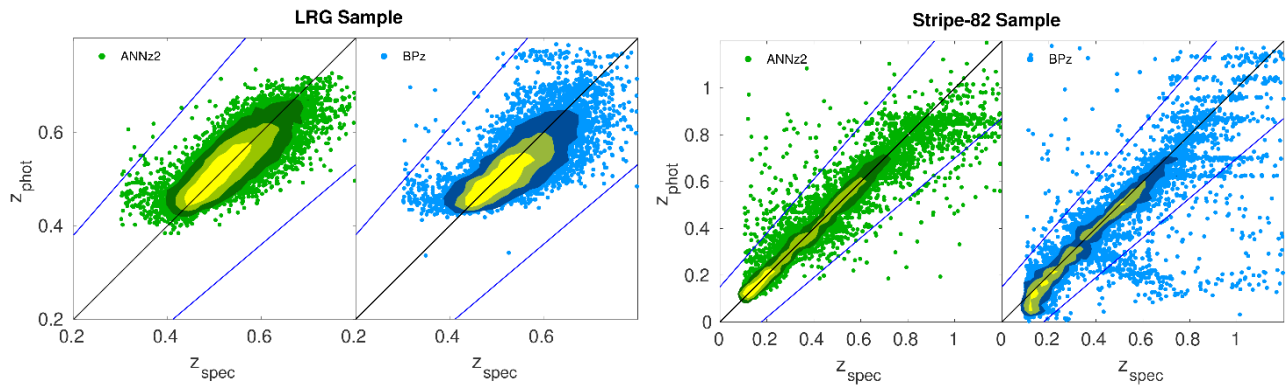

**Fig. 1**

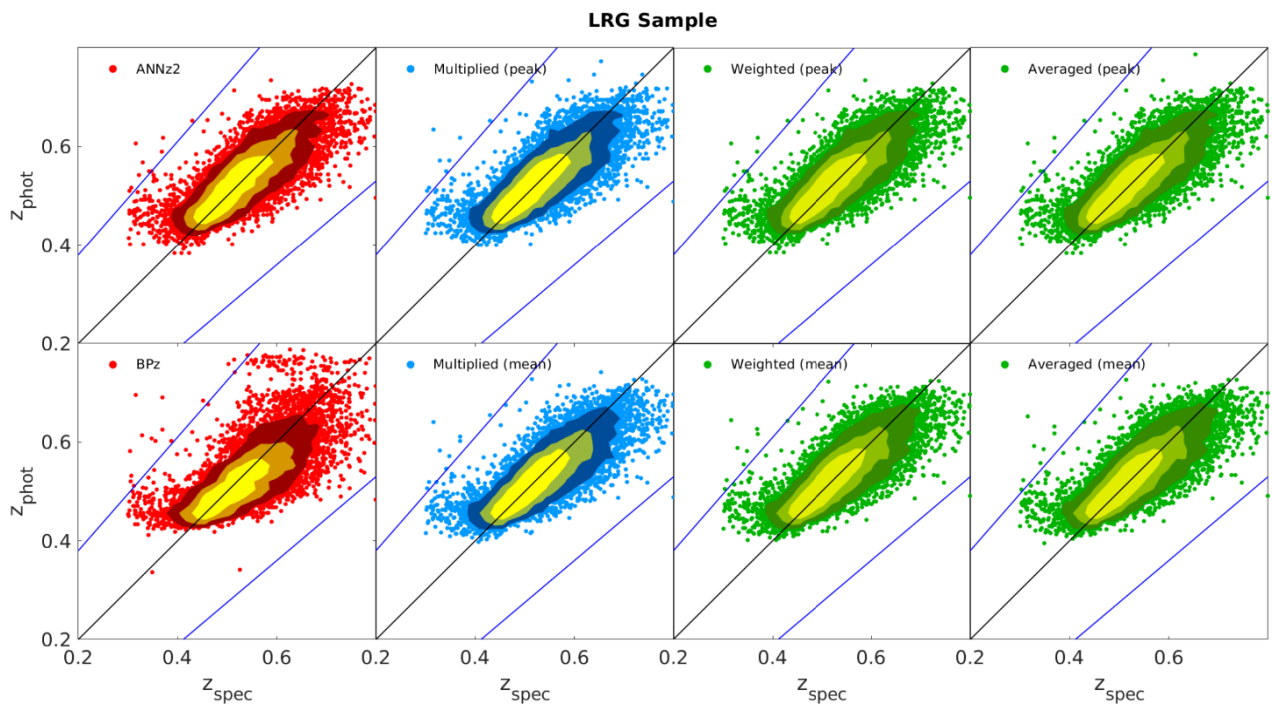

**Fig. 2.**

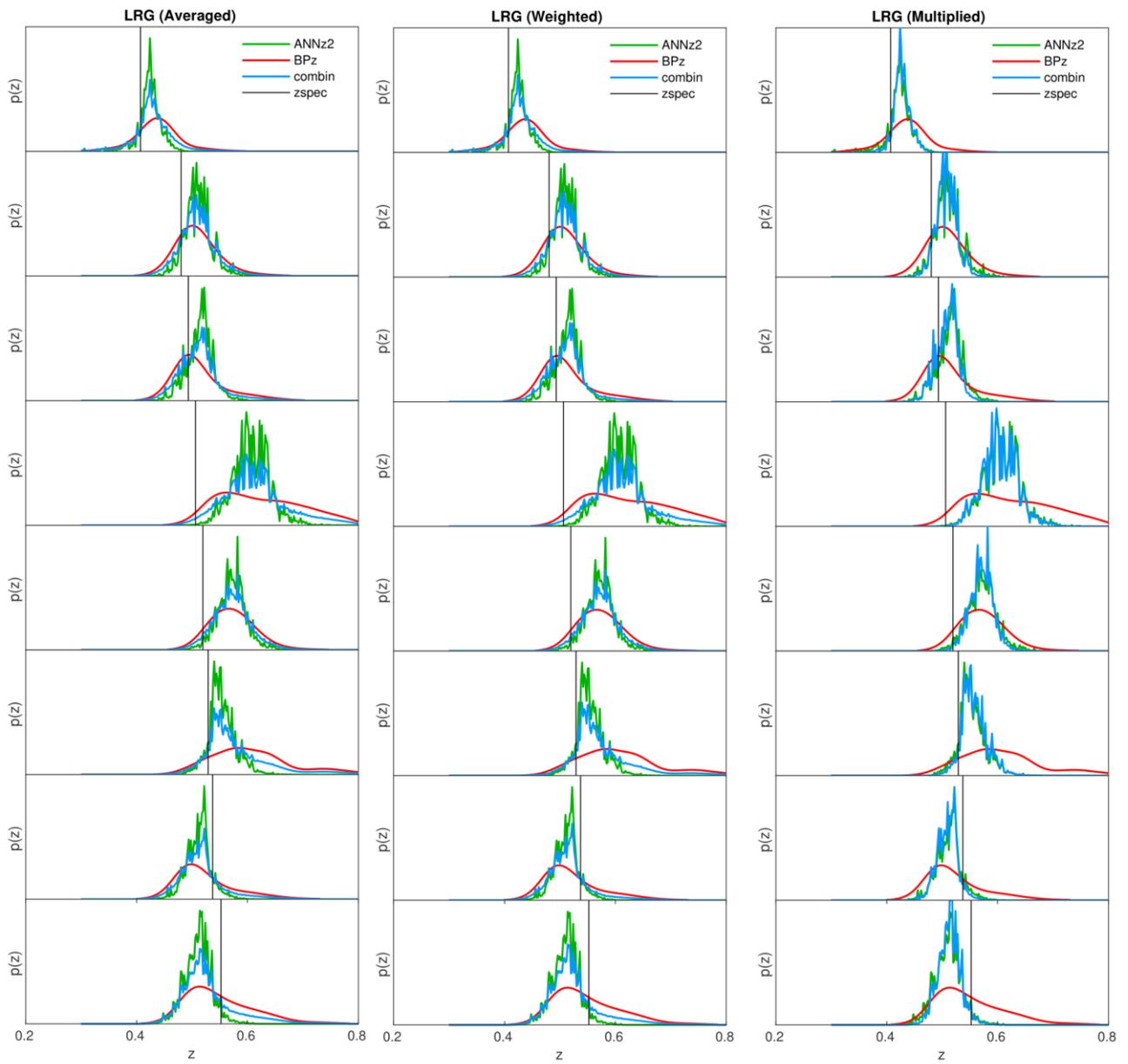

**Fig. 3.**

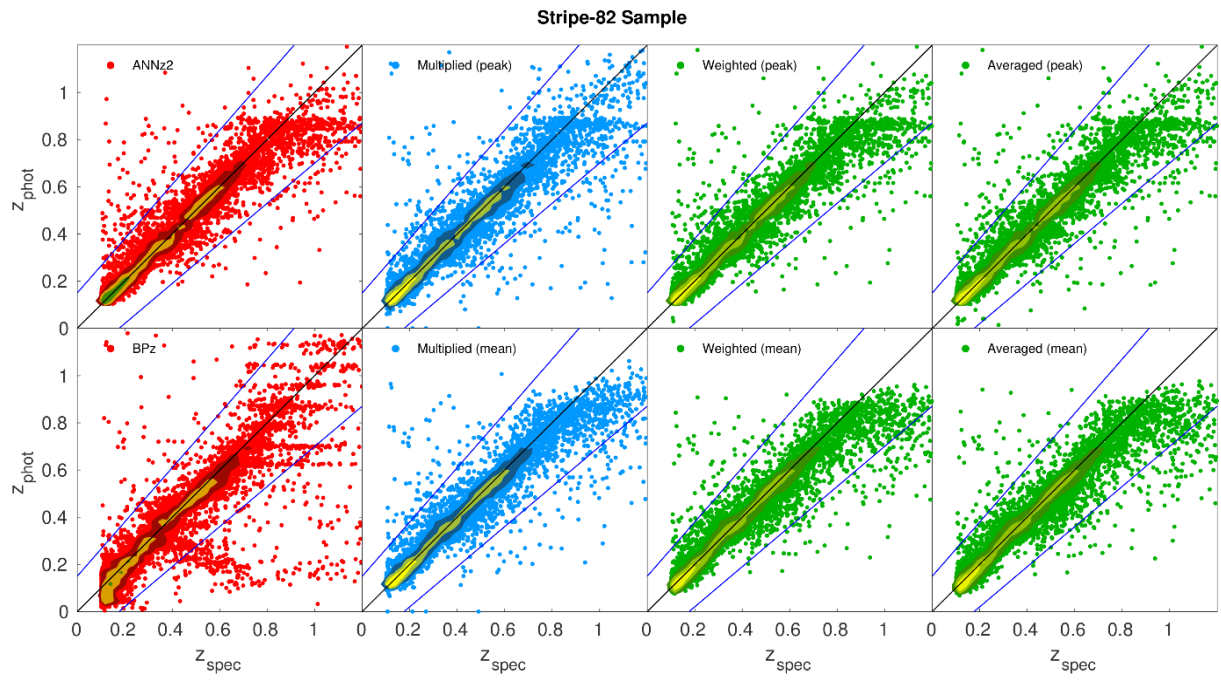

**Fig. 4.**

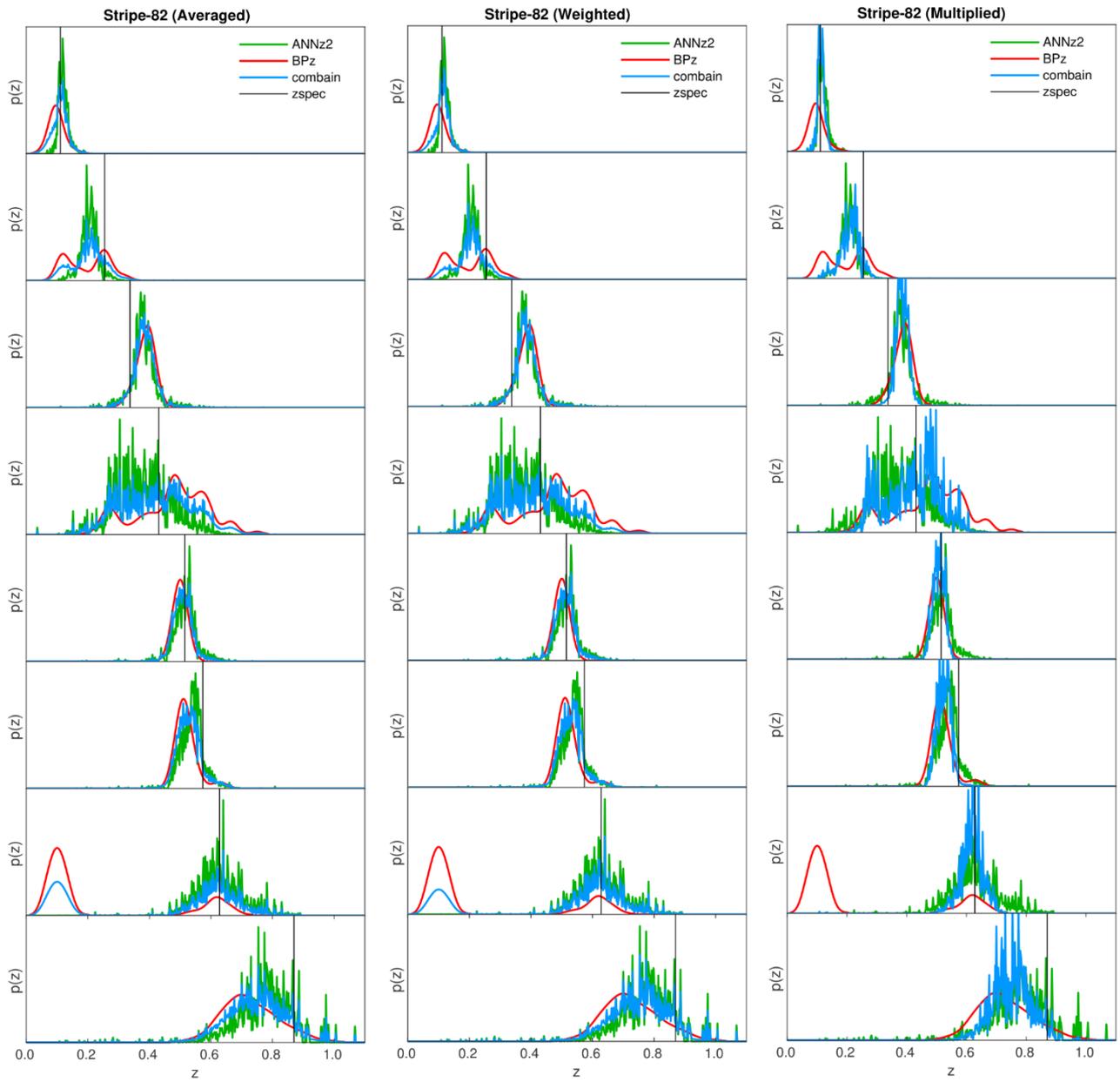

**Fig. 5.**